\documentclass{elsart}
\usepackage{epsf,fancyvrb}

\begin{document}%
\VerbatimFootnotes

\begin{frontmatter}
\begin{flushright}
SFB/CPP-06-15\\
TTP06-12\\
NIKHEF 06-002
\end{flushright}
\title{Extension of the functionality of the symbolic program FORM by 
external software}
\author[Karlsruhe]{M.~Tentyukov\thanksref{SFB}}
\author[NIKHEF]{and J.A.M.~Vermaseren}
\thanks[SFB]{On leave from Joint Institute for Nuclear Research,
141980 Dubna, Moscow Region, Russian Federation.}
\address[Karlsruhe]{
Institut f\"ur Theoretische Teilchenphysik,  Universit\"at Karlsruhe
D-76131 Karlsruhe, Germany
}
\address[NIKHEF]{
NIKHEF, Kruislaan 409, 1098 SJ, Amsterdam, The Netherlands
}

\begin{abstract}
We describe the implementation of facilities for the communication with 
external resources in the Symbolic Manipulation System FORM. This is done 
according to the POSIX standards defined for the UNIX operating system. We 
present a number of examples that illustrate the increased power due to 
these new capabilities.
\end{abstract}

\end{frontmatter}

\section{Introduction}

FORM \cite{VermaserenFORM,FORMdistribution} is a Symbolic Manipulation 
System (SMS) specialized to handle extremely large\footnote{The notion of 
large is of course time dependent. Currently we mean with this expressions 
that can have a number of terms of the order $\mathcal{O}(10^8)$ or more. This 
exceeds the current capabilities of Mathematica of Maple by a large 
factor.} expressions of many millions of terms in an efficient way. FORM 
allows mainly local operations on single terms, like replacing parts of a 
term or multiplying something to it (the ``locality principle'', see 
Sect.\ref{locality}). This special property of FORM allows it to deal with 
expressions that are much larger than the main memory available. It also 
enables parallelization~\cite{PARFORM}. Together these features make FORM a 
unique tool suitable for state of the art evaluations. It is widely used in 
Quantum Field Theory, where the calculation of very large sequences of 
algebraic expressions is required,
e.g.~\cite{FORMused1,% 
FORMused2,FORMused3,FORMused4,FORMused5,FORMused6,FORMused7,FORMused8,% 
FORMused9,FORMused10,FORMused11}.

Effective manipulation of large expressions requires that all algebraic 
instructions are applied to a big sequence of terms. The sheer size of the 
intermediate results prevents storage of more than a single version of an 
expression. Hence in FORM there is no benefit in ``interactivity'' and it 
is used in an non-interactive way. FORM provides a special programming 
language adapted for the manipulating of large sequences of algebraic terms 
and the user supplies programs written in this language. The major 
non-local operation which brings expressions to a standard form (the sort 
operation) makes the language powerful enough to program quite non-trivial 
algorithms; several FORM packages are widely used in multiloop 
calculations, e.g. MINCER \cite{MINCER} and MATAD \cite{MATAD}.

Because FORM programs mainly describe the treatment of individual terms, 
rather than that of complete expressions, and because there are few 
libraries for many popular algebraic operations, FORM is considered to be 
less user-friendly than for instance the Computer Algebra Systems (CAS) 
Mathematica or Maple. On the other hand, very large expressions and time 
consuming operations often need special rather than generic algorithms 
anyway and one cannot use the libraries in that case. There are however 
cases in which it is useful to have access to the facilities of other 
specialized software. That is why sometimes FORM is used in combination 
with other software systems which provide a more fluent control flow in 
conjunction with some specialized programs. An example is the combination 
of the programs QGRAF \cite{QGRAF}, Q2E, EXP \cite{EXP} and MATAD glued 
together with the help of the Make utility.

Several other systems use FORM as a sub-system for the evaluation of 
``hard'' expressions. The C-program DIANA \cite{Diana1,Diana2} designed as 
a master program for higher-order calculations uses FORM (see 
implementations e.g. \cite{FIRCLA,Aitalc} and references in 
\cite{Diana2}). The automatic calculation systems CompHEP
\cite{CompHEP,CompHEPMigrate}
 and GRACE \cite{GRACE} are 
currently migrating from REDUCE to FORM for analytical 
calculations. SANC \cite{SANC} is a recently developed world-wide network 
system which consists of SANC Servers and SANC Clients which uses FORM for 
analytical evaluations. Also FormCALC \cite{FormCALC} uses Mathematica as a 
user-friendly front-end to FORM.

As one can see, FORM is often used in co-operation with other software. 
This makes the improvement of the FORM abilities to communicate with other 
programs into an urgent problem.

The locality principle restricts the variety of possible algorithms or at 
least the way they should be programmed. It is impossible to avoid 
non-local operations completely. Hence FORM supports a number of non-local 
operations the most important of which is the sort operation. Another is 
the explicit bracketing of some common factor. By implementing a general 
interface for FORM to communicate with external programs, other non-local 
operations like polynomial factoring or the elimination of Greatest Common 
Divisors (GCD) between numerators and denominators can be delegated to 
other (less efficient but less restricted) programs. Recently, such an 
interface was elaborated. The interface permits FORM to run external 
programs and to be embedded in other applications. In the present paper we 
concentrate on the detailed description of the interface.

Of course, the systematic introduction to the FORM language is far
beyond the scope of this paper. There are different manuals and
tutorials devoted the FORM language, see the FORM 
homepage~\cite{FORMdistribution}.

After a very short introduction to the FORM language and the
mechanisms behind it in Sect.~\ref{forminternas}, we 
discuss different ways to communicate with the underlying operating
system (UNIX) available in FORM3.1 (the current FORM version),
Sect.~\ref{extworld}. The new mechanism available in the upcoming
version  FORM3.2 permitting a
dialog with an external program is presented in Sect.~\ref{async}.
The exact syntax is described in Sect.~\ref{exactsytntax}. A working
example of using the REDUCE~\cite{REDUCE} system from FORM is described and
discussed in Appendix~\ref{REDUCE}.

\section{General concepts}

\subsection{Common structure}
\label{forminternas}
\begin{figure}[ht]
\begin{center}
\epsfxsize=\linewidth \epsfbox{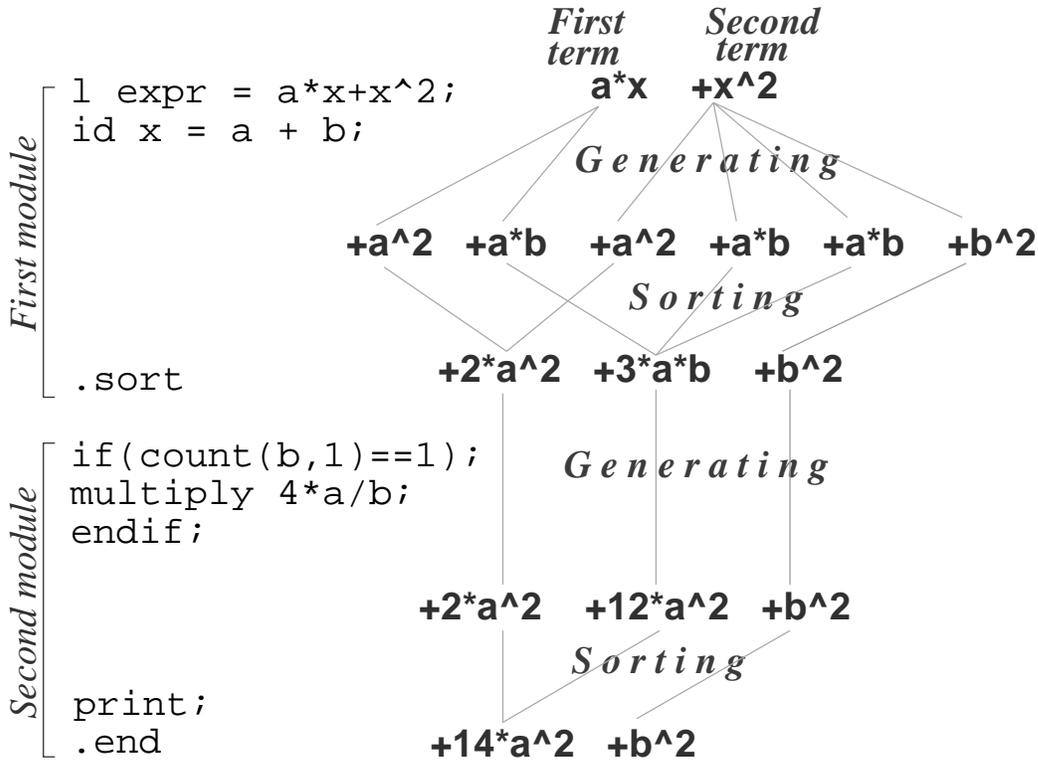}
\end{center}
\caption{
\label{example}
A fragment of a typical FORM program. In the first module the
expression $expr = ax+x^2$ is introduced, and then the substitution $x
\to a+b$ is performed. In the second module only terms in which the
degree of $b$ is exactly 1 are multiplied by $4a/b$ (there is only
one such a term in the expression).
}
\end{figure}

FORM is used non-interactively by
executing a program that contains several parts called {\em modules}.
The programs are executed module by module.
Modules are terminated
with ``dot''-instructions that cause the execution of the module, see
the example on the left of Fig.\ref{example}.

This  example consists of only two modules. There are two
``dot''-instructions: a {\tt .sort} and a {\tt .end}. In both cases the
result is sorted. {\tt .end} additionally terminates the program.

The execution of each module is divided into three steps:
\begin{itemize}
\item{\bf Compilation:} the input is translated into an internal 
representation.
\item{\bf Generating:} for each term of the input expressions the 
statements of the module are executed. This in general generates a lot of 
terms for each input term.
\item{\bf Sorting:} all the output terms that have been generated are 
sorted and equivalent terms are summed.
\end{itemize}

\subsection{The locality principle}
\label{locality}

Mostly FORM allows only local operations on single terms, like replacing 
parts of a term or multiplying something to it; non-local operations like
replacing a sum of two terms by another term are not allowed. 
We refer to this property as the {\em locality principle}:
all {\em explicit} algebraic operations are local.
Non-local operations are allowed only implicitly in the sorting
procedure at the end of the modules, when equivalent terms are summed up, 
and in some special references to the contents of brackets in the input 
expression. Together with a sophisticated pattern matcher, this at first 
strong limitation allows the formulation of general and efficient 
algorithms. The limitation to local operations makes it possible to handle 
expressions as ``streams'' of terms that can be read sequentially from 
memory or a file and processed independently, which allows dealing with 
expressions that are larger than the available main memory and in addition 
allows parallelism \cite{PARFORM}.

In principle non-local operations like polynomial factoring or GCD 
contraction could be  performed at the level of the preprocessor, but this 
means that we would like to design some new CAS in the frame of the FORM 
preprocessor. The more natural solution would be to delegate this problem 
to other (less efficient but less restricted) CAS. At this point we come to 
the problem of how to interact with the outside world.

\subsection{The control flow}

The generation of terms is performed by the algebraic processor applying 
the algebraic imperatives (the statements inside a module) to each term of 
all active expressions. Thus, the control flow immediately splits into many 
independent branches. The control flow can be manipulated by restricting 
the application of statements to those terms that fulfil some criteria as 
is done in the following \verb|if| statement
\begin{verbatim}
  if ( count(b,1) == 1 );
    multilpy 4*a/b;
  endif;
\end{verbatim}
The above means that we assign to each power of $b$ a degree of 1 and only 
terms in which the total degree is exactly 1 are multiplied by $4a/b$.
Control flow based on complete expressions is not possible at the algebraic 
processor level. But it is possible at the preprocessor level.

The preprocessor in FORM is a notably important ingredient. It coordinates 
the execution of the program and eventually calls the processor to perform 
the actual module execution. All the ``real'' control flow in FORM is 
performed by the preprocessor.
And, of course, interactivity is possible only at the preprocessor
level. The interactive mode ``Human -- FORM''
makes no sense because FORM usually deals with very big expressions,
and intermediate results that may take Gigabytes of storage are not 
readily open for inspection by the programmer. On the other hand the 
interaction ``FORM -- Another program'' often does make sense.
In principle, a realization of this could be delegated to the underlying
operating system. Of course, this introduces operating system
dependent code and should be used with great care.

\section {Interaction with the outside world}
\label{extworld}
\subsection{Running external programs in FORM3}
Let us first look at the existing mechanisms of running external programs 
in FORM3. The easiest model of interaction is just to start some operating 
system command and wait until it finishes. The preprocessor instruction
\begin{verbatim}
  #system systemcommand
\end{verbatim}
forces a system command to be executed by the operating system. The 
complete string (excluding initial blanks or tabs) is passed to the 
operating system. FORM will then wait until control is returned.

One evident drawback of this mechanism is that there is no  feedback from 
the system command. The only way to communicate is by means of the file 
system. It is quite clear that any nontrivial model would require that the 
output of the program is intercepted by FORM.

There is another preprocessor instruction, namely, \verb|#pipe|:
\begin{verbatim}
#pipe  systemcommand
\end{verbatim}
This forces a system command to be executed by the operating system. The 
complete string  (excluding initial blanks or tabs) is passed to the 
operating system. Next FORM will intercept the output of the command and 
read it as an input similar to the \verb|#include| instruction. Whenever 
output is produced FORM will take action, and it will wait when no output 
is ready. After the command has been finished, FORM will continue with the 
next line.

This preprocessor instruction is much more powerful than the former
one. But the communication channel established by the \verb|#pipe| 
instruction doesn't satisfy all our requirements. Evidently, this is mostly 
a {\em unidirectional} channel; FORM could read an output, but establishing 
a dialog with the running program is very tricky (see 
Appendix~\ref{PIPECM}).

Let us now try to formulate the requirements more precisely.

\subsection{Desired properties of the communication channel}

Even the simple \verb|#system| can be used for the extension of the FORM 
functionality. For example, consider an expression ``\verb|withGCD|'' which 
is a ratio of two polynomials in $d$:
\begin{equation}
\label{withGCD}
\frac{2d^4+3d^3-22d^2-13d+30}{d^3-11d+10}.
\end{equation}
Factoring both the numerator and the denominator we get
\begin{eqnarray}\nonumber
\frac{(d^2+d-10)(2d+3)(d-1)}{(d^2+d-10)(d-1)}&&
\end{eqnarray}
so after contracting the GCD we obtain
\begin{equation}
\label{noGCD}
2d+3.
\end{equation}
Suppose, there is some light-weight CAS which is able to do the GCD 
contraction. How could we proceed with the \verb|#system| instruction?

Consider the following simplified FORM program (for the full working
example see Appendix~\ref{SYSTEM}):
\begin{verbatim}
  #define cmd . . .
  Symbol d;
  Local withGCD = (2*d^4+3*d^3-22*d^2-13*d+30)/(d^3-11*d+10);
  #write <finput> "%E",withGCD
  #system cat finput | `cmd' > foutput
  Local noGCD =
  #include foutput
           ;
  print;
  .end
\end{verbatim}
Here we assume that the preprocessor variable \verb|`cmd'| contains the
command starting the external program.

First, we save the initial expression into the file ``finput'', then we run 
the external program performing the GCD contraction saving the result into 
another file ``foutput'' and at the end we read the result from the file. 
This works, and the result is given in the following simplified listing:
\begin{verbatim}
  . . .
    withGCD =
      30/(10-11*d+d^3)-13/(10-11*d+d^3)*d
     -22/(10-11*d+d^3)*d^2+3/(10-11*d+ 
      d^3)*d^3+2/(10-11*d+d^3)*d^4;

    noGCD =
      3 + 2*d;
\end{verbatim}
This is acceptable provided we need this only once or a few times but could 
be too time consuming for frequent usage like in a loop.

How can one reduce the overhead?

The use of intermediate files for input and output is, evidently,
a cost factor. Using the more powerful \verb|#pipe| instruction we would 
get rid of these files, e.g. by the following simplified FORM program:
\begin{verbatim}
  #define cmd . . .
  Symbol d;
  Local withGCD =(2*d^4+3*d^3-22*d^2-13*d+30)/(d^3-11*d+10);
  .sort
  $EXP=withGCD;
  .sort
  Local noGCD =
  #pipe echo "`$EXP'" | `cmd'
           ;
  Print;
  .end
\end{verbatim}
(for the full working example see Appendix~\ref{PIPE}).
Here instead of saving the input expression to a file we put it into 
the command line\footnote{Of course, the same trick we could use with 
the \verb|\#system| instruction; note however that this way the length of 
the input expression is restricted to a rather small value. This is more 
an illustration of the concept rather than a working approach.} and read 
the results back directly from the external command output.

But the overhead is not much less: for each evaluation we have to
start a new copy of the external program. This is because the
pipe establishes only a unidirectional channel with the external
program, so we are not able to organize a dialog with it.

In principle, the \verb|#pipe| preprocessor instruction can be used in 
order to establish a real dialog of FORM with the external program, see 
Appendix~\ref{PIPECM}. But this is quite tricky, extremely inconvenient and 
hardly applicable in practice. In order to extend the FORM functionality we 
need a really full duplex channel to the external program.

\subsection{Dialog with the external program}
\label{async}

Recently a full duplex interface has been implemented. The new preprocessor 
instruction \verb|#external| starts the system command opening an 
input-output channel for it. After the instruction \verb|#external| returns 
the control to the FORM program, external programs initiated by the 
instruction continue to run. The standard input and output are intercepted 
by FORM and not connected with any terminal device.

Generally speaking, the system command initiates several 
processes starting several external programs combined into the ``job''.
We refer to the job initiated by the instruction \verb|#external| as
the ``external command''.

The instruction \verb|#toexternal| is used to send some text to the running 
external command. Its syntax is similar to the \verb|#write| instruction.

The instruction \verb|#fromexternal| is used to read the text from the 
running external command. Its syntax is similar to the \verb|#include| 
instruction.

The (simplified, for the full example see Appendix~\ref{EXTERNAL}) variant 
of the above FORM program based on this new mechanism may look like:

\begin{verbatim}
  #define cmd . . .
  Symbol d;
  #external `cmd'

  Local withGCD =(2*d^4+3*d^3-22*d^2-13*d+30)/(d^3-11*d+10);
  .sort
  #toexternal "%E\n",withGCD
  Local noGCD =
  #fromexternal
          ;
  Print;
  .end
\end{verbatim}
Note that now we have a real dialog. 

After the instruction \verb|#external `cmd'| the control
returns to FORM 
while all the processes initiated by \verb|`cmd'| continue to run. The 
external command that is running is embedded in the FORM system and 
visible from the FORM program as a duplex channel. The instruction 
\verb|#toexternal| provides the external command with input data while the 
instruction \verb|#fromexternal| reads its resulting output.

It is interesting to compare the speed of all three programs. To minimize 
irrelevant overhead we repeat the corresponding evaluation 1000
times via a preprocessor \verb|#do| loop
and measure the wall-clock time. The results are as 
follows\footnote{The execution times refer to AMD Athlon XP 1800+ computer.}:

\begin{center}
\begin{tabular}{lrr}
\verb|#system| - based  &   506.8 & seconds \\
\verb|#pipe| - based  &    397.6 & seconds \\
\verb|#external| - based &   0.8   & seconds
\end{tabular}
\end{center}

\section{Dialog with the external command: detailed description}

\subsection{General remarks and short introduction}

The design of the full duplex channel is not so trivial; automatic 
bidirectional interaction can be dangerous due to possible 
deadlocks. This requires careful design and flexibility for all the 
communication primitives.

There are several modes of starting and finishing the external program. By 
default, the command is run in a subshell in a new 
session and in a new process group. Before FORM 
finishes (or by the \verb|#rmexternal| preprocessor instruction) the KILL 
signal is sent to the whole group. The preprocessor instruction 
\verb|#setexternalattr| could be used in order to change this default 
behavior.

Performing the \verb|#fromexternal| instruction FORM continues to read the 
output of the running external program until the external program outputs a 
{\em prompt}. The prompt is a line consisting of a given prompt string. By 
default, this is an empty string.  The prompt string can be changed by 
means of the preprocessor instruction \verb|#prompt|.

Several external commands can be run simultaneously. The external command 
that is started last becomes the {\em current} (active) one. Instructions 
\verb|#toexternal| and \verb|#fromexternal| deal with the current external 
command. At any time the current external command can be changed by the 
instruction \verb|#setexternal|.

This approach was chosen in order to reduce possible overhead. 
Alternatively, one could implement an additional argument to all primitives 
containing information about an identifier for each external command.

Special external channels (``pre-opened'') could be available
provided FORM was started by a parent process with some special
command line option. In that case FORM is able to communicate with the 
parent process by means of the same mechanism. This permits FORM to be 
embedded in other applications, see Appendix \ref{PREOPENED}.

An external command can be terminated by the instruction 
\verb|#rmexternal|.

\subsection{Exact syntax}
\label{exactsytntax}

\verb|#external ["prevar"] | systemcommand\\
starts the command in the background, connecting to its standard input and 
output. By default, the external command has no controlling terminal, the 
standard error stream is redirected to \verb|/dev/null| and the command is 
run in a subshell in a new session and in a new process group (see the preprocessor 
instruction \verb|#setexternalattr|).

The optional parameter ``prevar'' is the name of a preprocessor variable 
placed between double quotes. If it is present, the ``descriptor'' (small 
positive integer number) of the external command is stored into this 
variable and can be used for references to this external command (if there 
is more than one external command running simultaneously).

The external command that is started last becomes the ``current'' (active) 
external command.  All further instructions \verb|#fromexternal| and 
\verb|#toexternal| deal with the current external command.

\verb|#toexternal "formatstring" [,variables]|\\
sends the output to the current external command. The semantics of the 
\verb|"formatstring"| and the \verb|[,variables]| is the same as for the 
\verb|#write| instruction, except for the trailing end-of-line symbol. In 
contrast to the \verb|#write| instruction, the \verb|#toexternal| 
instruction does not append any new line symbol to the end of its output.

\verb!#fromexternal[+|-] ["[$]varname",[maxlength]]!\\
appends the output of the current external command to the FORM program. The 
semantics differ depending on the optional arguments. After the external 
command sends the prompt, FORM will continue with a next line after the 
line containing the \verb|#fromexternal| instruction. The prompt string is 
not appended. The optional + or - sign after the name has influence on
the listing of the content.

\verb!#fromexternal[+|-]!\\
The semantics is similar to the \verb|#include| 
instruction but folders are not supported. 

\verb!#fromexternal[+|-] "[$]varname"!\\
is used to read the text from the running external 
command into the preprocessor variable \verb|varname|, or
into the dollar variable \verb|$varname| if the name of the variable
starts with the dollar sign ``\$''.

\verb!#fromexternal[+|-] "[$]varname" maxlength!\\ is used to read
the text from the running external command into the preprocessor (or
dollar) variable \verb|varname|. Only the first \verb|maxlength|
characters are stored.
	
The prompt is a line consisting of a single prompt string. By default, this 
is an empty string.  The prompt can be changed by means of the instruction 
\verb|#prompt|:

\verb|#prompt [newprompt]|\\
Sets a new prompt for the current external command (if present) and all 
further (newly started) external commands.

If \verb|newprompt| is an empty string, the default prompt (an empty line) 
will be used.

\verb|#setexternal n|\\
sets the ``current'' external command. The instructions \verb|#toexternal| 
and \verb|#fromexternal| deal with the current external command.  The 
integer number \verb:n: must be the descriptor of a running external 
command.

\verb|#rmexternal [n]|\\
terminates an external command. The integer number \verb|n| must be either 
the descriptor of a running external command, or 0.

If \verb|n| is 0, then all external programs will be terminated.

If \verb|n| is not specified, the current external command will be terminated.

The action of this instruction depends on the attributes of the external 
channel (see the preprocessor instruction \verb|#setexternalattr|). By 
default, the instruction closes the commands' IO channels, sends a KILL 
signal to every process in its process group and waits for the external 
command to be finished.

\verb|#setexternalattr| list\_of\_attributes\\
sets attributes for {\em newly started} external commands. Already
running external commands are not affected. The list of attributes
is a comma separated list of pairs attribute=value, e.g.:
\begin{verbatim}
  #setexternalattr shell=noshell,kill=9,killall=false
\end{verbatim}
Possible attributes are:

\begin{itemize}
\item
\verb|kill|  specifies the signal to be sent to the external command 
either before the termination of the FORM program or by the preprocessor 
instruction \verb|#rmexternal|. By default this is 9 (SIGKILL). Number 0 
means that no signal will be sent.
\item
\verb|killall| Indicates whether the kill signal will be sent to the whole 
group or only to the initial process. Possible values are ``\verb|true|'' 
and ``\verb|false|''. By default, the kill signal will be sent to the
whole group.
\item
\verb|daemon| Indicates whether the command should be ``daemonized'', i.e. 
the initial process will be passed to the init process and will belong
to the new process group in the new session.  
Possible values are ``\verb|true|'' and ``\verb|false|''. By default, 
``\verb|true|''.
\item
\verb|shell| specifies which shell is used to run a command. By default 
this is ``\verb|/bin/sh -c|''.  If set \verb|shell=noshell|, the command 
will be stared by the instruction \verb|#external| directly but not in a 
subshell, so the command should be a name of the executable file rather 
than a system command. The instruction \verb|#external| will duplicate the 
actions of the shell in searching for an executable file if the specified 
file name does not contain a slash (/) character.  The search path is the 
path specified in the environment by the PATH variable.  If this variable 
isn't specified, the default path ``\verb|:/bin:/usr/bin|''
is used.
\item
\verb|stderr| specifies a file to redirect the standard error stream to. 
By default it is ``\verb|/dev/null|''.
\end{itemize}
Only attributes that are explicitly mentioned are changed, all others remain 
unchanged. Note, changing attributes should be done with care. For example,
\begin{verbatim}
  #setexternalattr daemon=false,kill=9,killall=true
\end{verbatim}
starts a command in the subshell within the current process group.
The instruction \verb|rmexternal| sends the KILL signal to the whole
group, which means that also FORM itself will be killed.

The example:
\begin{Verbatim}[numbers=left,numbersep=8pt]
  symbol a,b;

  #external "n1" cat -u

  #external "n2" cat -u

  *  cat simply repeats its input. The default prompt is an
  *  empty line. So we use "\n\n" here -- one "\n" is to finish
  *  the line, and the next "\n" is the prompt:
  #toexternal "(a+b)^2\n\n"

  #setexternal `n1'
  *  For this channel the prompt will be "READY\n":
  #toexternal "(a+b)^3\nREADY\n"

  #setexternal `n2'
  *  Set the default prompt:
  #prompt
  Local aPLUSbTO2=
  #fromexternal
         ;

  #setexternal `n1'
  #prompt READY
  Local aPLUSbTO3=
  #fromexternal
         ;

  #rmexternal `n1'
  #rmexternal `n2'

  Print;
  .end
\end{Verbatim}

Two external channels are opened in lines 3 and 5. The UNIX utility 
``\verb|cat|'' simply repeats its input.
The option
``\verb|-u|'' is used to prevent the output buffering. The option 
is ignored by the GNU \verb|cat| utility but is
mandatory for non-GNU versions of \verb|cat|. 

After line 5 the current external channel is `\verb|n2|'.
The default prompt is an empty line so in line 10 ``\verb|\n\n|'' is used -- one
``\verb|\n|'' is to finish the line, and the next ``\verb|\n|'' is the prompt.

Line 12 switches the current channel to `\verb|n1|'. For this channel
the prompt will be ``\verb|READY|'', see line 24, hence the expression is
finished by ``\verb|\nREADY\n|''.

Line 16 switches to the `\verb|n2|' external channel and line 18 sets
the default prompt (which is extra in this example since the default
prompt was not changed up to now).

Results (just a literal repetition of the sent expressions) are read
in lines 20 and 26.

Lines 29 and 30 close the external channels.

\section{Conclusion}
An interface like the one described above can be used in order to extend 
FORM in the spirit of a component model where the term ``component model'' 
means that a software system is built from ``components''. Components are 
high level aggregations of smaller software pieces and provide a ``black 
box'' building block approach to software construction (see, e.g., 
\cite{Szyperski}).

The communication is performed only via standard input and output streams 
without any special protocol, i.e., the input is expected to follow the 
FORM syntax conventions. But this is not restrictive since the 
communication could easily be established through an intermediate gateway 
program (also called filter) which performs a translation between the 
external programs and FORM (see the discussion in Appendix~\ref{FERMAT} and 
\ref{REDUCE}).

There are two important points:

\begin{itemize}
\item
The interface is implemented at the preprocessor level. This means that the 
structure of the FORM algebraic processor is not touched and all the FORM 
advantages persist.
\item
It would also allow us to implement a model of parallelization, in which 
several external components are running simultaneously at the preprocessor 
level while the FORM algebraic processor continues to work on some huge 
expression.
\end{itemize}

The technique described is applied directly to FORM swallowing the
external program, or to embedding FORM in other applications.  However,
this mechanism is quite general and can be used also in order to build
FORM in the client-server architecture.

The idea is that FORM can use the ``\verb|#external...|'' 
instruction in order to start a client or a server. The client (or the
server) then establishes a connection with an external application and
uses its standard input and output in order to communicate with FORM.

The corresponding library (``formlink'') is under development and will
be available from the FORM distribution site \cite{FORMdistribution}
in the near future.

\section*{Acknowledgements}

We want to thank K. Chetyrkin, H.-M. Staudenmaier and M. Steinhauser for 
interesting discussions. We would also like to thank A. Grozin for 
advises and useful remarks concerning REDUCE and C. Sturm for 
continuous testing of the software in a real application. The work of M.T. 
has been supported by the Deutsche Forschungsgemeinschaft in 
Sonderforschungsbereich/Transregio 9 ``Computational Particle Physics''. 
The work of J.V. has been part of the research program of the Dutch 
Foundation for Fundamental Research of Matter (FOM).

\appendix
\section*{Appendix}

\section{Choosing an external CAS}
\label{FERMAT}

The technique we presented was developed and tested in an experimental FORM 
version used in multiloop calculations \cite{STURM} applying the Laporta 
algorithm \cite{LAPORTA}. The implementation requires some operations on 
polynomials with integer/rational coefficients.

There are surprisingly few decent programs available for this. 
It should be a program performing some basic operations on 
polynomials, and it should be completely stateless, i.e. allow to handle input
expressions independent of each other. Ideally, this program should be:
\begin{enumerate}
\item
   fast enough;
\item
   able to handle very large expressions;
\item
   available on most hardware/software platforms;
\item
   open source.
\end{enumerate}
For the above example we found only two suitable programs, 
REDUCE~\cite{REDUCE} and FERMAT~\cite{FERMAT}.

REDUCE is fast enough, it is an open source system (but {\em not} a free 
one!) but on 32-bit platforms it is rather restrictive, 
and there is no IA64 (Intel 
Itanium) - based version of REDUCE at the moment. Appendix~\ref{REDUCE} 
contains an example of a fully functional embedding of REDUCE in FORM.

In our examples\footnote{These examples are based on ref.~\cite{STURM}}
we selected FERMAT, ``conjugated'' with the 
\verb|#external| -based FORM interface by a gateway program which performs 
a syntax translation and proper masking \cite{AIR} in order to isolate FORM 
from long expressions treated by FERMAT.

The gateway starts FERMAT, reads input from stdin, passes expressions in 
the variable d (indicating the dimension of space-time in our example), 
which FORM collected into the function acc, to FERMAT, reads an answer from 
FERMAT, stores the answer into an internal table, contracts it to the 
string ``\verb|dd(#)|'', where ``\verb|#|'' is the order number of the 
expression produced by FERMAT, and outputs the result to the stdout.
Here is a very short introduction to the ideas behind it.
For example, the expression
\begin{verbatim}
  some string + acc((d+1)/(d-1)+d) + another string
\end{verbatim}
will be converted to
\begin{verbatim}
  some string + dd(1) + another string
\end{verbatim}
where \verb|dd(1)| is equal to \verb|(d^2+1)/(d-1)|. It is stored 
internally and may be extracted by means of the command @v, see below. All 
lines started with the character @ are commands for the gateway. We show 
some of them (just a few ones):
\begin{itemize}
\item
@f0 - do not filter the content of acc() through FERMAT;\\
@f1 - filter the content of acc() through FERMAT (default);
\item
@e0 - reduce all rational functions in d in acc() to dd(\#) (default);
@e1 - expand all dd(\#) to corresponding
rational functions (not only in acc()!);
\item
@vsome text @(\#) another text - repeat the line after @v,
 substituting @(\#) by the content of the corresponding dd-variable.
 Example: ``@vid dd(1) = @(1);''
 will result in ``id dd(1) = (d\^{}2+1)/(d-1);''
\item
@sfilename - all stored variables will be saved into the file
 ``filename'',
  one variable per line, in the order as they were stored (i.e., dd(5)
 will
  occupy the line number 5);\\
@rfilename - all lines from the file ``filename'' (without trailing
 '\verb|\n|')  will be loaded into the internal table.
\end{itemize}

FERMAT is not an open source program; it is available only for Macintosh, 
Sun SPARC, and Windows and Linux IA32-based platforms (and even not for all 
of them) but it is able to handle very large expressions even on 32-bit 
platforms. FERMAT is approx. 20 times faster than the Mathematica function 
``Together[]'' and a little bit faster than REDUCE. It is used for 
performing the operations on polynomials arising in coefficient functions; 
we need only basic polynomial arithmetic (which, in principle, FORM 
provides itself, but still in an experimental and far from optimized way) 
and the only non-local operation we really need from FERMAT is a GCD 
contraction.

FERMAT is an interpreter providing a complete programming language. Most of 
its commands begin with the symbol \verb|&| and have the syntax 
\verb|&<symbol>| or \verb|&(<symbol>=<value>)|, as in \verb|&q| for 
quitting the program. There are also 
some other classes of commands
and the terminal I/O commands ! and ?. Entered commands are separated by a 
new line or by a semicolon.

Entered arithmetic expressions are reduced to some unique form and printed 
to the terminal. Two arithmetic modes are possible in FERMAT, rational 
arithmetic and modular. A session with FERMAT starts in rational mode (see 
example below). All arithmetic is that of rational numbers and the GCD of a 
numerator and a denominator is contracted. The user may change the ground 
field by converting it into a polynomial ring entering the ``adjoin 
polynomial'' command, \verb|&J|. After that, the user may enter any 
polynomial sums and ratios and FERMAT will bring them into a unique form of 
a rational function (``quolynomial'').

In the following example of the FERMAT session  the character ``\verb|>|'' 
is the FERMAT prompt, characters after the prompt are typed by the user, 
and strings without the prompt are printed by FERMAT. We do not show some 
of the housekeeping information printed by FERMAT:
\begin{Verbatim}[numbers=left,numbersep=8pt]
  >21/14
   3 / 2;  or  1.5000000000000
  >&d
   Enter display size. 
  >0
  >21/14
   3 / 2
  >&(J=d) 
  >(d^2 - 4)/(d+2)
    d - 2
  >&(J=x)
  >(x^2 + 2*x*d +d^2)^2/(x+d)
    x^3 + (3d)x^2 + (3d^2)x + d^3
\end{Verbatim}
The command \verb|&d| is used in order to change the number of blocks of 8 
significant digits to the right of the decimal point displayed by the 
interpreter when a non-integer is displayed. If it is 0, this feature is 
disabled.

In line 8 we change the ground field (rationals) to the ring of polynomials 
in $d$, and after line number 11 the ground ring becomes the ring of 
polynomials in the two variables $d$ and $x$.

We describe briefly some other FERMAT commands which are used in the rest 
of the paper.

\verb|&(M=' ')| Change the prompt to just an empty line.

\verb|&(t=0)| Switch off the output of information about the time.

\verb|&U| Toggle switch to display long integers and polynomials in the 
style of other computer algebra systems (Maple).

\section{Examples of running external programs}

\label{RUNEXTCMD}

Here we compare all techniques that were available in FORM3 with the new 
techniques available in FORM 3.2. We provide examples of FORM programs but 
these programs are only simplified conceptual (but working) examples of how 
this could be done.  General comments:

\begin{itemize}
\item
As an external CAS,
in Appendix~\ref{RUNEXTCMD} and \ref{PIPECM}
we use the program FERMAT, see 
Appendix~\ref{FERMAT}. We assume that the executable file is available from 
the current directory as ``./fermat''.
\item
For simplicity, we do not pay any attention to the line lengths. In 
reality, some intermediate gateway program (script) should be used in order 
to make a proper formatting. All the FORM programs in the examples below 
are started with the  statement ``\verb|format 254;|''. This means, the 
examples may not work if the total transferred length exceeds 254 
characters.
\item
In order to compare the speed of the examples in all the versions, we 
repeat the evaluation 1000 times by means of a preprocessor \verb|#do| 
loop. However, in the example in Appendix~\ref{PIPECM} the loop is 
performed by a ``while'' loop in the shell.
\item
The common structure of all the examples is the same. We create the local 
expression ``withGCD'' which is Expr.~(\ref{withGCD}), and subsequently 
repeat 1000 times the process of filtering the expression through FERMAT 
and putting the resulting Expr.~(\ref{noGCD}) back into FORM.
\end{itemize}

\subsection{Based on the {\tt \#system} instruction}
\label{SYSTEM}

The following FORM program exploits the \verb|#system| preprocessor 
instruction to start FERMAT. The information is transferred by conventional 
files.

\begin{Verbatim}[numbers=left,numbersep=8pt]
  Off statistics;
  Format 254;
  #define cmd "./fermat |grep LE=|sed s/^.\*LE=//"
  #define prolog "&U; &(J=d); !(\'LE=\',"
  #define epilog "); &q"
  Symbol d;
  Local withGCD = (2*d^4+3*d^3-22*d^2-13*d+30)/(d^3-11*d+10);
  .sort
  #do i = 1,1000
    #write <finput> "`prolog' %E `epilog'",withGCD
    #system cat finput | `cmd' > foutput
    #remove <finput>
    Local noGCD =
    #include foutput
             ;
    Print;
    .sort
    Drop noGCD;
    .sort
  #enddo
  .end
\end{Verbatim}
First the expression \verb|withGCD| is saved to the file ``finput'' (line 
10), then FERMAT is started (line 11). The contents of the input file is 
passed to the FERMAT standard input and the answer is redirected to the 
file ``foutput''. Note that the file \verb|finput| must be removed (line 
12) or else in each iteration the \verb|#write| instruction will append the 
input expression to the file.

After FERMAT has been started, the content of the file ``finput'' is sent to 
it:
\begin{verbatim}
  &U; &(J=d); !('LE=',(2*d^4+3*d^3-22*d^2-13*d+30)/
                                            (d^3-11*d+10)); &q
\end{verbatim}
FERMAT prints a lot of text, and one line with the answer:
\begin{verbatim}
  LE=  2*d+3
\end{verbatim}
The answer is selected by the filter 
\begin{verbatim}
  grep LE=|sed s/^.\*LE=//
\end{verbatim}
and redirected to the file ``foutput''.

The program is placed into the file ``system.frm'' and the following 
command is executed:
\begin{verbatim}
  ( time form system.frm ) > system.res 2>&1 
\end{verbatim}
The result is redirected to the file system.res. Here are several lines
from the end of the file system.res:
\begin{verbatim}
     noGCD =
        3 + 2*d;

      .end

  real  8m26.815s
  user  5m4.010s
  sys   0m24.250s
\end{verbatim}

\subsection{Based on the {\tt \#pipe} instruction}
\label{PIPE}

The following FORM program uses the \verb|#pipe| preprocessor instruction 
to start FERMAT:
\begin{Verbatim}[numbers=left,numbersep=8pt]
  Off statistics;
  #define cmd "|./fermat |grep LE=|sed s/^.\*LE=//"
  #define prolog "\&U\; \&\(J=d\)\; \!\(\'LE=\',"
  #define  epilog "\)\; \&q"
  Symbol d;
  Local withGCD =(2*d^4+3*d^3-22*d^2-13*d+30)/(d^3-11*d+10);
  .sort
  $EXP=withGCD;
  .sort
  #do i=1,1000
    Local noGCD =
    #pipe echo `prolog' "`$EXP'" `epilog' `cmd' 
             ;
    Print;
    .sort
    Drop noGCD;
    .sort
  #enddo
  .end
\end{Verbatim}
The initial expression is passed to FERMAT after the initialization 
commands which use the command ``echo'' directly from the command line. The 
answer is read from the FERMAT standard output intercepted by FORM in the 
instruction \verb|#pipe|.

The local expression \verb|withGCD| defined on line 6 is passed to the 
preprocessor on line 12 by means of a dollar variable \verb|$EXP| 
initialized on line 8.

The program is placed into the file ``pipe.frm'' and the following command 
is executed:
\begin{verbatim}
  ( time form pipe.frm ) > pipe.res 2>&1 
\end{verbatim}
The result is redirected to the file pipe.res. We show several lines
at the end of the file pipe.res:
\begin{verbatim}
     noGCD =
        3 + 2*d;
  
      .end

  real  6m37.593s
  user  5m3.360s
  sys   0m23.750s
\end{verbatim}

\subsection{Based on the {\tt \#external} instruction}
The following FORM program uses the \verb|#external| preprocessor 
instruction to start FERMAT:
\label{EXTERNAL}
\begin{Verbatim}[numbers=left,numbersep=8pt]
  Off statistics;
  Format 250;
  #define cmd "./fermat"
  #define init "&d\n0\n&(M=\' \')\n&(t=0)\n&U\n&(J=d)\n"
  Symbol d;
  #external `cmd'
  #toexternal "`init'"
  #do i = 1,19
    #fromexternal "tmp"
  #enddo
  Local withGCD =(2*d^4+3*d^3-22*d^2-13*d+30)/(d^3-11*d+10);
  .sort
  #do i=1,1000
    #toexternal "%E\n",withGCD
    #fromexternal "tmp"
    Local noGCD =
    #fromexternal
             ;
    Print;
    .sort
    Drop noGCD;
    .sort
  #enddo
  .end
\end{Verbatim}

After FERMAT has been started on line 6 the initialization string is sent 
in line 7. FERMAT answers with 19 lines of text which are read into the 
preprocessor variable ``\verb|tmp|'' and subsequently ignored (lines 8-10). 
After the expression has been sent (line 14), FERMAT produces an empty line 
(which is read in line 15 and ignored) and the output expression is 
assigned to the local expression ``\verb|noGCD|'' (lines 16-18).

The program is placed in the file ``extern.frm'' and the following 
command is executed:
\begin{verbatim}
  ( time form extern.frm ) > extern.res 2>&1 
\end{verbatim}
The result is redirected to the file extern.res, of which we show several 
lines at the end:
\begin{verbatim}
     noGCD =
        3 + 2*d;
  
      .end
  
  real  0m0.799s
  user  0m0.160s
  sys   0m0.010s
\end{verbatim}

\section{ Duplex channel with the {\tt \#pipe} instruction}
\label{PIPECM}

A ``real'' dialog between FORM and some external program is possible only 
if the full duplex channel is established between FORM and the external 
program. In principle, this could be done based on the \verb|#pipe| 
instruction using the following trick (we don't claim this to be the only 
way in which it can be done).

\begin{figure}[ht]
\begin{center}
\ \epsfxsize=.7\linewidth \epsfbox{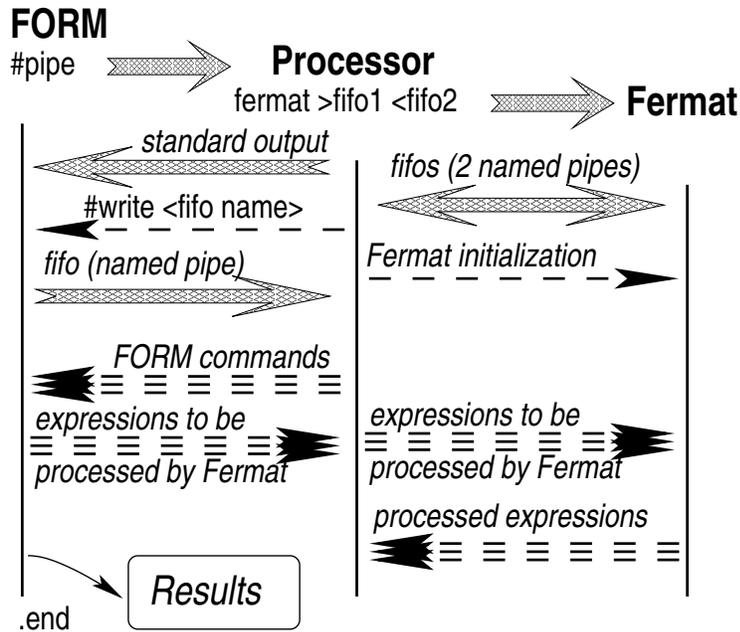} \\
\end{center}
\caption{
\label{process}
The ``main'' processor program is started by FORM from the trivial
program and performs the main control. After initialization, the main
program sends a flow of FORM commands to FORM getting back a flow
of expressions to be processed by FERMAT. This flow is passed to FERMAT,
and the flow of processed expressions is analyzed and transformed back 
into a flow of expressions transferred to FORM.
}
\end{figure}

The ``main'' processor program could be not FORM but something started by
FORM via the \verb|#pipe| instruction. FORM is only waiting for commands
from the main processor, performs these commands and returns results
back by means of some sockets or named pipes, see Fig.~\ref{process}. 
The FORM program itself is rather trivial:
\begin{verbatim}
  Off statistics;
  Format 250;
  #pipe processor.sh
  .end
\end{verbatim}
The main functionality is contained by the shell script ``processor.sh'':
\begin{Verbatim}[numbers=left,numbersep=8pt]
  #!/bin/bash
  rm -f tofermat fromfermat fromform 2> /dev/null
  mkfifo tofermat fromfermat fromform
  ./fermat < tofermat > fromfermat &
  exec 3 > tofermat
  exec 4 < fromfermat
  echo -e "&d\n0\n&(M=' ')\n&(t=0)\n&U\n&(J=d)" > &3
  echo "!('PROMPT')" > &3
  while [ "$REPLY" != "PROMPT" ] ; do
  read -u 4
  done
  read -u 4; read -u 4
  echo '#write < fromform > "PROMPT"'
  echo 'Symbol d;'
  exec 6 < fromform
  read -u 6
  echo "Local withGCD =(2*d^4+3*d^3-22*d^2-13*d+30)\
                                               /(d^3-11*d+10);"
  echo ".sort"
  i=0
  while [ $i -lt 1000 ] ; do
  let i++;
  echo '#write < fromform > "%E",withGCD'
  echo
  read -u 6
  echo "$REPLY" > &3
  read -u 4; read -u 4;
  echo "Local noGCD = ${REPLY};"
  read -u 4
  echo 'Print;'
  echo '.sort'
  echo 'Drop noGCD;'
  echo '.sort'
  done
  rm -f tofermat fromfermat fromform
\end{Verbatim}
This file must be executable, which is achieved by entering the command\\
``\verb|chmod +x ./processor.sh|''\\
after the file has been created.
The script is the ``real'' main program and uses FORM for some algebraic
evaluations.

The communication between FORM and FERMAT is established by the script 
using several FIFOs (named pipes).  These pipes are created at lines 2 and 
3. The named pipe ``tofermat'' will be used in order to send data to FERMAT 
while the named pipe ``fromfermat'' will be used in order to receive data 
from FERMAT. The named pipe ``fromform'' will be used in order to get data 
from FORM while data to FORM  are sent by the script via the standard 
output.

First, the script starts FERMAT (line 4) intercepting the output by opening 
corresponding FIFOs (lines 5 and 6). Then it initializes FERMAT (line 7) 
and reads all the unimportant stuff produced by FERMAT (lines 8 -- 12).

Then the script asks FORM to create the local expression \verb|withGCD|, 
reads the result (line 25) and sends it to FERMAT (line 26). The FERMAT 
answer is, again, passed to FORM (line 28) and then the script asks FORM to 
print the result (line 30).

The FORM program is placed into the file ``wrapper.frm'' and the following 
command is executed:
\begin{verbatim}
  ( time form wrapper.frm ) > process.res 2>&1 
\end{verbatim}
The result is redirected to the file process.res. Here are several lines
at the end of the file process.res:
\begin{verbatim}
     noGCD =
        3 + 2*d;
  
      drop noGCD;
      .sort
      .end
  
  real  0m1.340s
  user  0m0.500s
  sys   0m0.060s
\end{verbatim}

Note that logically the  main processor program (the script
``processor.sh'') appears to be dependent on the ``stub''-like wrapper
FORM program ``wrapper.frm''. The ``main'' processor program should
really parse the input in order to decide what should be sent to FORM
and which data should be passed to FERMAT. Practically, the processor
program is a new CAS using FORM to evaluate large-scale expressions,
and FERMAT is used for non-local algebraic operations. This is rather 
similar to the use of Perl scripts in which alternatingly FORM and Maple 
are used to solve problems in Mathematics~\cite{jansanders}.

\section{Embedding FORM in other applications}
\label{PREOPENED}

The external channel instructions permit FORM to swallow an external 
program. The same mechanism can be used in order to {\em embed} FORM in 
other applications.

There is a possibility to start FORM from another program providing
one (or more) communication channels (see below). These channels will be 
visible from a FORM program as ``pre-opened'' external channels existing 
after FORM starts. There is no need to open them with the 
``\verb|#external|'' instruction.
In this case, the preprocessor variable ``\verb|PIPES_|''
 is defined
and is equal to the total number of the pre-opened external channels.
Pre-opened external channel descriptors are contained in the preprocessor
variables ``\verb|PIPE1_|'', ``\verb|PIPE2_|'', etc.
For example, if `\verb|PIPES_|' is 3 then there are 3 
pre-opened external channels with the descriptors `\verb|PIPE1_|',
`\verb|PIPE2_|' and `\verb|PIPE3_|' so e.g. the following instruction could
be used:
\begin{verbatim}
#setexternal `PIPE2_'
\end{verbatim}
without 
\begin{verbatim}
#external "PIPE2_"
\end{verbatim}

The external channel attributes make no sense for the pre-opened channel 
(see the preprocessor instruction \verb|#setexternalattr|).
Formally, they are as follows:
\begin{verbatim}
   kill=0,
   killall=false,
   daemon=false,
   stderr=/dev/tty,
   shell=noshell
\end{verbatim}

In order to activate the pre-opened external channels, the parent
application must follow some standards. Here we describe a low-level 
protocol, the corresponding C-interface is available
from the FORM distribution site \cite{FORMdistribution}.

Before starting FORM, the parent application must create one or more pairs 
of pipes\footnote{A pipe is a pair of file descriptors, one is for reading 
and another is for writing. See ``man 2 pipe''.}. The read-only descriptor 
of the first pipe in the pair and the write-only descriptor of the second 
pipe must 
be passed to FORM as an argument of a command line option
``\verb|-pipe|'' in ASCII decimal format. The argument of the option
is a comma-separated list of pairs ``\verb|r#,w#|'' where
``\verb|r#|'' is a read-only descriptor and ``\verb|w#|'' is a
write-only descriptor. For example, to start FORM with two pre-opened
external channels the parent application has to create first four
pipes. Lets us suppose the first pipe was created with the descriptors 5
and 6, the second pipe has the descriptors 7 and 8, the third pipe has
the descriptors 9 and 10 and the fourth pipe has the descriptors 11 and 12.
The descriptors 5 and 8 will be used by FORM as the input and the output for
the first pre-opened external channel while the descriptors 9 and 12
will be used by FORM as the input and the output for
the second pre-opened external channel.

Then the parent application must start FORM with the following 
command line option:
\begin{verbatim}
   -pipe 5,8,9,12
\end{verbatim}

Upon startup, FORM sends its PID (the Process Identifier) in ASCII
decimal format with an appended 
newline character to the descriptor 8 and then FORM
will wait for the answer from the descriptor 5.
The answer must be two comma-separated integers in ASCII decimal format
followed by a newline character. The first integer corresponds
to the FORM PID while the second one is the parent process PID.
If the answer is not obtained after some timeout, or if it is not
correct (i.e. it is not a list of two integers or the first integer is
not the FORM PID) then FORM fails. If everything is correct, FORM creates
the pre-opened channel and puts its descriptor in the preprocessor
variable ``\verb|PIPE1_|''.

Then FORM processes the second pair of arguments, ``\verb|9,12|''.

After all pairs have been processed FORM creates the preprocessor variable
``\verb|PIPES_|'' and puts into this variable the total number of created
pre-opened external channels.

The order of processing the pairs of numbers in the argument is fixed 
exactly as it was described above i.e. from the left to the right.

\section{Using REDUCE from FORM}
\label{REDUCE}

The presented technique can also be used to embed the REDUCE system in 
FORM.

Transferring data to REDUCE we could ask FORM to use the output
format to be compatible with the REDUCE one and the REDUCE output is
essentially understandable by FORM although some minor syntax translation
should be done. Hence we do need a gateway. The following simple shell
script could be used as a gateway:
\begin{Verbatim}[numbers=left,numbersep=8pt]
  #!/bin/sh
  reduce=/export/pc/reduce
  MACHINE=linux
  lisp=psl
  IMAGE=$reduce/lisp/$lisp/$MACHINE/red/reduce.img
  export MACHINE
  export reduce
  export lisp

  echo 'off int$' > reduceinit
  echo 'lisp setq(promptstring!*,"")$' >> reduceinit
  echo 'lisp setq(promptexp!*'",'! )$" >> reduceinit
  echo 'off nat$'  >> reduceinit
  echo 'on gcd$'   >> reduceinit
  echo 'on ezgcd$' >> reduceinit
  echo '1;' >> reduceinit

  REDUCE="$reduce/lisp/psl/$MACHINE/psl/bpsl -td 128000000 \
                                                    -f $IMAGE"
  OUTF1='sed -u  s/\$/\nP\n/g'
  OUTF2='sed -u /^$/d'
  cat -u reduceinit -|$REDUCE|$OUTF1|$OUTF2
  rm -f reduceinit
\end{Verbatim}
The shell variable \verb|reduce| defined on line 2 is the REDUCE 
root (dependent of the REDUCE installation!). 
The variables \verb|MACHINE|, \verb|reduce| and \verb|lisp| should 
be exported in order for REDUCE to work properly (lines 3-8).

Next the script prepares the initialization commands. These commands are 
placed into the file \verb|reduceinput| (lines 10-16) which will be deleted 
afterwards (line 23). The contents of the file are copied to REDUCE using 
the standard input, see the command \verb|cat| (line 22). The argument 
``\verb|-|'' means that the standard input will be appended ({\em after} 
the contents of the initialization file \verb|reduceinput|). The option 
``\verb|-u|'' is ignored by the GNU \verb|cat| utility  but is mandatory 
for non-GNU versions of \verb|cat|.

The problem here is that most utilities attached to something different 
than the terminal perform so called ``block buffering'', i.e., the data are 
passed to the standard output only when the internal buffer is full or when 
the end-of-file condition is reached. Evidently, we need a different 
behavior: the data should be passed every time the string is ready. This 
is so called ``line buffering'', or just unbuffered output. There are two 
UNIX utilities supporting this mode, namely \verb|cat| and \verb|sed|. The 
mode should be switched on explicitly by the \verb|-u| option (the GNU 
version of \verb|cat| does this by default, and the option \verb|-u| is 
just ignored).

The prompt is set to ``\verb|P|''. The end of the REDUCE output expression 
is the character ``\verb|$|'', which the output filter ``\verb|OUTF1|'' 
(line 20) replaces by a line that contains only the string ``\verb|P|''. 
The ``\verb|OUTF2|'' filter on line 21 removes empty lines. The 
initialization commands are: \hfill \\
\verb|off int$| -- do not remember a history; \hfill \\
\verb|lisp setq(promptstring!*"")$|, \verb|lisp setqpromptexp!*,'! )$| -- 
set the prompt string to be a new line character; \hfill \\
\verb|off nat$| -- switch on a ``flat'' output; \hfill \\
\verb|on gcd$| -- contract a greatest common divisor; \hfill \\
\verb|on ezgcd$| -- use the EZ GCD algorithm (Extended Zassenhaus GCD, 
\hfill \\
http://www.uni-koeln.de/REDUCE/3.6/doc/reduce/node120.html). \hfill \\
The input \verb|1;| is added in order to get the prompt string ``\verb|P|''.

Let us suppose the script is placed into the executable file 
``\verb|runreduce|'' available in the current directory. The following FORM 
program, named \hfill \\
``\verb|usereduce.frm|'' demonstrates how to embed 
REDUCE in FORM:
\begin{Verbatim}[numbers=left,numbersep=8pt]
  Symbol d;
  Format Reduce;
  #prompt P
  #external ./runreduce
  #fromexternal "tmp" 1
  Local withGCD =(2*d^4+3*d^3-22*d^2-13*d+30)/(d^3-11*d+10);
  .sort
  #toexternal "%e\n",withGCD
  Local noGCD =
  #fromexternal
           ;
  Print;
  .sort
  #toexternal "(%E)*(%E)*(%E)";\n,withGCD,withGCD,withGCD
  Local noGCD1 =
  #fromexternal
           ;
  Print;
  .sort
  #toexternal "(%E)^20;\n",withGCD
  Local noGCD2 =
  #fromexternal
           ;
  Print;
  .end
\end{Verbatim}

The external channel to REDUCE is created on line 4. REDUCE responds
to initialization by several lines which are read on line 5 and
ignored.

Manipulating the output format of FORM is sometimes not so
convenient. The ``reduce'' format option is not guaranteed to be
perfect. Using a gateway, we could translate the FORM default output
into the REDUCE syntax ``on the fly''.

Let us suppose, the following script ``\verb|runreduce1|'' is 
available in the current directory:
\begin{Verbatim}[numbers=left,numbersep=8pt]
  #!/bin/sh
  reduce=/export/pc/reduce
  MACHINE=linux
  lisp=psl
  IMAGE=$reduce/lisp/$lisp/$MACHINE/red/reduce.img
  export MACHINE
  export reduce
  export lisp

  echo 'off int$' > reduceinit
  echo 'lisp setq(promptstring!*,"")$' >> reduceinit
  echo 'lisp setq(promptexp!*'",'! )$" >> reduceinit
  echo 'off nat$'  >> reduceinit
  echo 'on gcd$'   >> reduceinit
  echo 'on ezgcd$' >> reduceinit
  echo '1;' >> reduceinit

  REDUCE="$reduce/lisp/psl/$MACHINE/psl/bpsl -td 128000000 \
                                                     -f $IMAGE"
  INF1='sed -u s/\^-[0123456789]*/&Pr/g'
  INF2='sed -u s/\-[0123456789]*Pr/(&)/g'
  INF3='sed -u -e s/Pr//g -e s/\^/**/g'
  OUTF0='sed -u  s/\*\*/\^/g'
  OUTF1='sed -u  s/\$/\nP\n/g'
  OUTF2='sed -u /^$/d'
  cat -u reduceinit \
               -|$INF1|$INF2|$INF3|$REDUCE|$OUTF0|$OUTF1|$OUTF2
  rm -f reduceinit
\end{Verbatim}
This file must be executable, which it can be made by entering the 
command \hfill \\
``\verb|chmod +x ./runreduce1|'' \hfill \\
after the file has been created.

The script differs from the former script ``\verb|runreduce|'' by a 
sequence of input and output filters, making the complete syntax 
translation.

Input filters are defined in lines 20-22. The problem here is that
REDUCE does not understand expressions like \verb|a^-1|; it
needs at least \verb|a^(-1)| instead of it. The standard symbol for
the power operation in REDUCE is ``\verb|**|'' so all ``\verb|^|'' are
replaced by ``\verb|**|'' (though this is not necessary since REDUCE
understans the symbol `\verb|^|'' in the input correctly).

The ``\verb|INF1|'' filter translates the ``bare'' negative powers like 
``\verb|a^-1|'' into the form ``\verb|a^-1Pr|'', the filter ``\verb|INF2|'' 
converts it into an expression like ``\verb|a^(-1Pr)|'' and the filter 
``\verb|INF3|'' finally replaces the power operation symbol ``\verb|^|'' by 
the REDUCE-standard ``\verb|**|'' and removes the auxiliary tag 
``\verb|Pr|''.

The output filter ``\verb|OUTF0|'' (line 23) translates the REDUCE power 
operation ``\verb|**|'' to the FORM notation ``\verb|^|''. In principle, 
this is not necessary since FORM understands the REDUCE-like input 
notation.

The ``\verb|usereduce.frm|'' should now be modified: 
line 2 (``\verb|Format Reduce;|'') must be removed, and line 4 
must be changed into the following one:
\begin{verbatim}
  #external ./runreduce1
\end{verbatim}

Note, REDUCE will produce some strange line breaks when the numbers
become too big, hence a simple gateway like \verb|runreduce1| is not 
completely suitable. If, for instance, line number 20 of the file 
``\verb|usereduce.frm|'' is changed into
\begin{verbatim}
  #toexternal "(%E)^100;\n",withGCD
\end{verbatim}
FORM fails reading the data produced by REDUCE.

The problem is that REDUCE could break the line just
before the symbol ``\verb|*|'' and FORM ignores the next
line assuming this is a comment. The comment character could
be changed but in general it is more reliable
to get the answer without linebreaks at all.

Unfortunately, there is no standard UNIX utility permitting
unbuffered operations on the end-of-line symbol. The only way here is
to write some small program, e.g., the following Perl script
placed in the file ``\verb|smallfilter.pl|'':
\begin{Verbatim}[numbers=left,numbersep=8pt]
  #!/usr/bin/perl
  $|=1;
  while(<>){
    $_ =~ s/\*\*/^/g;
    $_ =~ s/\n//g;
    $_ =~ s/\$/\nP\n/g;
    print; 
  }

\end{Verbatim}
This file must be executable, which it can be made by entering the command 
\hfill \\
``\verb|chmod +x ./smallfilter.pl|'' \hfill \\
after the file has been created.

This simple script transfers the REDUCE power operation expression 
(\verb|**|) to 
the FORM one (\verb|^|), adds the prompt to the end of the output 
expression and glues all lines into a single string. 

The command in line 2 switches off buffering.

Line 4 
of ``\verb|smallfilter.pl|'' is 
equivalent to the filter ``\verb|OUTF0|'' in line 23 of the script
``\verb|runreduce1|''. Line 6 
of ``\verb|smallfilter.pl|'' is 
equivalent to the filter ``\verb|OUTF1|'' in line 24 of the script
``\verb|runreduce1|'', and line 5
of ``\verb|smallfilter.pl|'' extends the functionality of the filter
``\verb|OUTF2|'' in line 25 of the script \hfill \\
``\verb|runreduce1|'' just gluing a whole portion of the output
into a single line.

Using this script, line (26,27) in the file ``\verb|runreduce1|'' must be
written as
\begin{verbatim}
  cat -u reduceinit -|$INF1|$INF2|$INF3|$REDUCE|./smallfilter.pl
\end{verbatim}
and the lines 23-25 should be removed.
The resulting gateway is usable without any limitations.

Of course, it is much better to write the gateway as a whole program 
from the beginning. The corresponding program can be written in Perl:
\begin{Verbatim}[numbers=left,numbersep=8pt]
#!/usr/bin/perl
use IPC::Open2;
$|=1;

$reduce="/export/pc/reduce";
$MACHINE="linux";
$lisp="psl";
$IMAGE="$reduce/lisp/$lisp/$MACHINE/red/reduce.img";
$ENV{"MACHINE"}=$MACHINE;
$ENV{"reduce"}=$reduce;
$ENV{"lisp"}=$lisp;

$cmd="$reduce/lisp/psl/$MACHINE/psl/bpsl";
$arg=" -td 128000000 -f $IMAGE";
open2(*R, *W, $cmd . $arg);

print W  "off int\$\n";
print W "lisp setq(promptstring!*,\"\")\$\n";
print W "lisp setq(promptexp!*,'! )\$\n";
print W "off nat\$\n";
print W "on gcd\$\n";
print W "on ezgcd\$\n";
print W "1;\n";
$outp=<R> until $outp eq "1\$\n" ;

while(<>){
   $_ =~ s/\^\-([0-9]+)/^(-$1)/g;
   $_ =~ s/\^/**/g;
   print W $_;
   $RANSW=m/;/;
   while($RANSW){
      $outp=<R>;
      $outp =~ s/\*\*/^/g;
      $outp =~ s/\n//g;
      $outp =~ s/(\$)/\nP\n/g;
      print $outp;
      $RANSW=($1 ne '$');
   }
}
print W "bye;\n";
<R>;<R>;
close(R);
close(W);
\end{Verbatim}

REDUCE is started (line 15) by means of a standard IPC::Open2 module
(line 2). Line 3 switches off buffering. Lines 5 - 8 set necessary 
paths\footnote{Remember, in order to run the script, the path to 
the REDUCE root \hfill \\
(``\verb|/export/pc/reduce|'') should be changed according to the
local REDUCE installation.}
and some environment variables which are exported on lines 9 - 11. 
Initialization commands (lines 17 - 23) are the same as discussed
above (lines 10 - 16 of the file ``\verb|runreduce|''). In contrast to
shell scripts, the REDUCE response is read directly from the script and
ignored (line 24).

Lines 26 - 39 contain the main loop.

The script reads line by line from the standard input (in the
conditional expression of the \verb|while| loop, line 26) into the
default variable \verb|$_|. After some syntax translation (lines 27
and 28) the translated line is sent to REDUCE (line 29). Line 27
corresponds to the input filters \verb|INF1| and \verb|INF2| and line 28
corresponds to the filter \verb|INF3| of the script
``\verb|runreduce1|'' (see lines 20 - 22 of the file
``\verb|runreduce1|''). In principle, the line 28 is not needed since
REDUCE understands the power symbol ``\verb|^|'' in the input expression.

Once the script detects a semicolon in an input line (lines 30, 31) 
it starts to read an
answer from REDUCE, lines 31 - 38, performing a (generally speaking,
unnecessary) syntax translation from REDUCE to FORM (line 33). All
newline characters are removed (line 34) and the REDUCE
end-of-expression character ``\verb|$|'' is converted to the prompt
``\verb|P|''. If such a conversion occurs, the variable
``\verb|RANSW|'' becomes \verb|false| (line 37) and the script continues to read
the standard input.

The script should be placed into an executable file ``\verb|runreduce.pl|''
and can be used with the FORM program ``\verb|usereduce.frm|'', where 
line 4 must be replaced by:
\begin{verbatim}
  #external ./runreduce.pl
\end{verbatim}
and line number 2 must be removed. Also line 5 must be removed since
now the gateway removes all the messages produced by REDUCE at
initialization, see line 24.

All these examples are available on-line from the FORM distribution 
site \cite{FORMdistribution}.

\end{document}